\documentclass[useAMS,usecolumn,usenatbib,usegraphicx]{mn2e}

\title[The ALMA Patchy Deep Survey]{The ALMA Patchy Deep Survey: A blind search for [C\,{\Large II}] emitters at $z\sim 4.5$}
\author[Y. Matsuda et al.]{
\parbox[t]{\textwidth}{\vspace{-1cm}
Y. Matsuda,$^{\! 1,2}$\thanks{E-mail: yuichi.matsuda@nao.ac.jp} T. Nagao,$^{\! 3}$ D. Iono,$^{\! 1,2}$ B. Hatsukade,$^{\! 1}$ K. Kohno,$^{\! 4,5}$ Y. Tamura,$^{\! 4}$ Y. Yamaguchi,$^{\! 4}$ and I. Shimizu$^{6}$}\\\\
$^{1}$National Astronomical Observatory of Japan, Osawa 2-21-1, Mitaka, Tokyo 181-8588, Japan\\
$^{2}$Graduate University for Advanced Studies (SOKENDAI), Osawa 2-21-1, Mitaka, Tokyo 181-8588, Japan\\
$^{3}$Research Center for Space and Cosmic Evolution, Ehime University, Bunkyo-cho 2-5, Matsuyama, Ehime 790-8577, Japan\\
$^{4}$Institute of Astronomy, The University of Tokyo, 2-21-1 Osawa, Mitaka, Tokyo 181-0015, Japan\\
$^{5}$Research Center for the Early Universe, The University of Tokyo, 7-3-1 Hongo, Bunkyo, Tokyo 113-0033, Japan\\
$^{6}$Department of Astronomy, The University of Tokyo, 7-3-1 Hongo, Bunkyo, Tokyo 113-0033, Japan\\
}
\begin{document}

\date{Accepted ... ; Received ... ; in original form ...}

\pagerange{\pageref{firstpage}--\pageref{lastpage}} \pubyear{2012}

\maketitle

\label{firstpage}

\begin{abstract}

We present a result of a blind search for [C\,{\sevensize II}]\,158\,$\mu$m emitters at $z\sim 4.5$ using ALMA Cycle\,0 archival data.  We collected extra-galactic data covering at $330-360$\,GHz (band\,7) from 8 Cycle\,0 projects from which initial results have been already published.  The total number of fields is 243 and the total on-source exposure time is 19.2\,hours.  We searched for line emitters in continuum-subtracted data cubes with spectral resolutions of $\sim$50, 100, 300 and 500\,km\,s$^{-1}$.  We could not detect any new line emitters above a 6-$\sigma$ significance level.  This result provides upper limits to the [C\,{\sevensize II}] luminosity function at $z\sim 4.5$ over $L_{\rm [C\,{\sevensize II}]} \sim 10^8 - 10^{10} L_{\odot}$ or star formation rate, $SFR \sim 10 - 1000$~M$_{^\odot}$yr$^{-1}$.  These limits are at least 2 orders of magnitude larger than the [C\,{\sevensize II}] luminosity functions expected from the $z \sim 4$ UV luminosity function or from numerical simulation.  However, this study demonstrates that we would be able to better constrain the [C\,{\sevensize II}] luminosity function and to investigate possible contributions from dusty galaxies to the cosmic star-formation rate density by collecting Cycle\,1+2 archival data as the ALMA Patchy Deep Survey.

\end{abstract}


\begin{table*}
 \centering
 \begin{minipage}{177mm}
  \caption{Summary of published archival ALMA Cycle\,0 band\,7 data used for the emission-line search}
  \begin{tabular}{@{}cccccccccc@{}}
  \hline
   Project ID (Region$^a$)  & Targets & \# of Fields & Antennas & $t_{\rm exp}^b$ & RMS$^c$ & Synthesised beam  & Bandwidth & Ref.$^d$\\
           & & & & (min) & (mJy) & (FWHM) & (GHz) & \\
 \hline
 2011.0.00020.S (EA) & NGC1614 & 1 & $16$ & 51 & $0.65-0.70$ & $1''.3-1''.5$ & 7.5 & 1\\
 2011.0.00039.S (EU) & SBS0335$-$052 & 1 & $24$ & 243 &  $0.14-0.18$ & $0''.5-0'' .7$ & 7.5 & 2\\
 2011.0.00097.S (NA) & COSMOS & 114 & $17-24$ & 279 & $0.67-2.3$ & $0''.4-1''.0$ & 8 & 3\\
 2011.0.00101.S (EA) & GRBs & 2 & $17-20$ & 91 & $0.39-0.48$ & $0''.9-1'' .6$ &  8 & 4\\
 2011.0.00108.S (EA) & NGC1097 & 1 & $14-15$ & 59 & $0.35-0.42$ & $1''.2-1'' .5$ &  7.5 & 5\\
 2011.0.00208.S (EU) & NGC1433 & 1 & $19$ & 101 & $0.33-0.39$ & $0''.4-0'' .6$ &  7.5 & 6\\
 2011.0.00294.S (EU) & ECDFS & 120$^e$ & $12-15$ & 244 & $1.1-5.4$ & $1''.0-3''.8$  & 8 & 7\\
 2011.0.00467.S (EA) & VV114 & 3 & $18-20$ & 85 &  $0.27-0.30$ & $0''.4-0''.5$ &  7.5 & 8\\
\hline
\end{tabular}
$^a$The ALMA Regions, EA: East Asia, EU: Europe, NA: North America.\\
$^b$The total on-source time (minutes).\\
$^c$The 1-$\sigma$ sensitivity at 300 km s$^{-1}$ spectral resolution.\\
$^d$1: \citet{2013AJ....146...47I}, 2: \citet{2014A&A...561A..49H}, 3: \citet{2014ApJ...783...84S}, 4: \citet{2012ApJ...761L..32W}, 5: \citet{2013ApJ...770L..27F}, \citet{2013PASJ...65..100I}, 6: \citet{2013A&A...558A.124C}, 7: \citet{2012MNRAS.427..520C}, \citet{2012MNRAS.427.1066S, 2014MNRAS.438.1267S}, \citet{2013MNRAS.432....2K}, \citet{2013MNRAS.431L..88H}, \citet{2013ApJ...768...91H}, \citet{2013ApJ...778..179W}, \citet{2014ApJ...780..115D}, 8: \citet{2013PASJ...65L...7I}, \citet{2013ApJ...777..126S}, \citet{2014ApJ...781L..39T}\\
$^e$The data of two $z=4.4$ SMGs are excluded.\\
\end{minipage}
\end{table*}

\section{Introduction}

The [C\,{\sevensize II}] $^2$P$_{3/2} \rightarrow ^2$P$_{1/2}$ fine-structure transition at 1900.5469\,GHz (157.74\,$\mu$m) is a dominant coolant of the inter-stellar medium (ISM) in galaxies \citep[e.g.,][]{1985ApJ...291..755C, 1999ApJ...511..721C, 2013ApJ...774...68D, 2014ApJ...788L..17D}.  The [C\,{\sevensize II}] line is one of the brightest lines at far-infrared (FIR) and is expected to be a tracer of star formation rate in local to distant galaxies \citep[e.g.,][]{2011MNRAS.414L..95S, 2011MNRAS.416.2712D, 2014arXiv1402.4075D, 2012ApJ...755..171S, 2014arXiv1405.5759S}. 

Since the emission line can hold up to 1\% of the bolometric luminosity of a galaxy, it is also very bright, allowing 'blind' redshifts to be measured.  Indeed, ALMA observations of sub-millimetre galaxies (SMGs) serendipitously detected [C\,{\sevensize II}] emission from two SMGs at $z=4.4$ \citep{2012MNRAS.427.1066S}.  They obtained the first constraint to the [C\,{\sevensize II}] luminosity function ($L_{\rm [C\,{\sevensize II}]} > 10^9 L_{\odot}$) at $z=4.4$, using the original survey area with APEX/LABOCA \citep[LESS,][]{2009ApJ...707.1201W} and the [C\,{\sevensize II}] redshift coverage in the 7.5\,GHz bandwidth.  However, their constraint was only lower limit because their targets are continuum-selected objects and they could miss continuum-faint [C\,{\sevensize II}] emitters.  

The [C\,{\sevensize II}] luminosity function may be a useful tool to estimate the cosmic star-formation rate density at $z>4$, and the evolution of star-formation rate density can provide tests of galaxy formation models \citep[e.g.,][]{2013ApJ...762L..31B}.  However, at $z>4$, the contribution from dusty galaxies to the cosmic star formation rate density is still uncertain \citep[e.g.,][]{2013A&A...554A..70B, 2014ApJ...780...75D, 2014arXiv1403.0007M, 2014MNRAS.438.1267S}.  In order to obtain better constraint to the [C\,{\sevensize II}] luminosity function at $z>4$, we carried out a blind search for [C\,{\sevensize II}] emitters at $z\sim 4.5$ using ALMA Cycle\,0 archival data.  Throughout this Letter, we adopt a cosmology with $H_0 = 72$~km~s$^{-1}$~Mpc$^{-1}$, $\Omega_m = 0.27$, and $\Omega_{\Lambda}=0.73$.

\begin{figure}
\includegraphics[width=85mm]{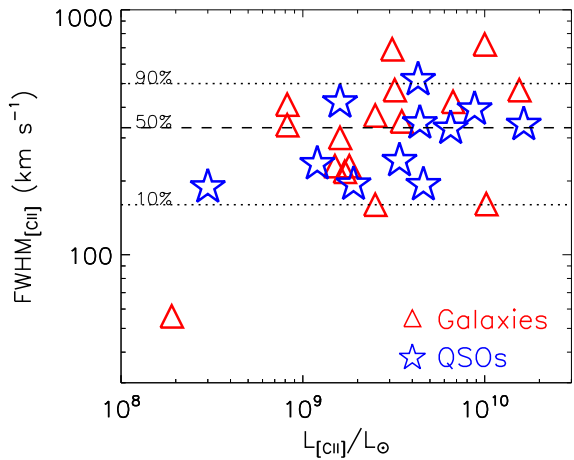}
  \caption{Distribution of [C\,{\sevensize II}] velocity width (FWHM) is shown as a function of the [C\,{\sevensize II}] luminosity ($L_{\rm [C\,{\sevensize II}]}$) for a sample of [C\,{\sevensize II}]-detected galaxies/QSOs at $z>4$ in literature.  The [C\,{\sevensize II}] velocity width ranges from $\sim 50 - 700$\,km\,s$^{-1}$.  The dashed line shows the median [C\,{\sevensize II}] velocity width of FWHM\,$\sim$\,330\,km\,s$^{-1}$.  The dotted lines represent the 10 and 90 percentiles.  We use this plot to motivate the velocity resolution of the four sets of data cubes used in our line emitter search.}
\end{figure}

\section[]{Data and Results}

In ALMA science archive, we searched Cycle\,0 Projects covering $330-360$\,GHz (in band\,7), which corresponds to the [C\,{\sevensize II}] redshift range of $z=4.28-4.76$.  We collected 8 extra-galactic projects from which initial results have been already published.  We only used the published data because we were able to know the data quality from the papers before we downloaded the data from the archive.  We summarise the properties of the data sets in Table\,1.  The data contain very deep single pointing and shallow multiple pointings.  The data with 7.5\,GHz bandwidth were taken with spectral mode (FDM mode, the spectral resolution is 0.98\,MHz or $\sim$0.85\,km\,s$^{-1}$) and the data with 8\,GHz bandwidth were taken with continuum mode (TDM mode, the spectral resolution is 31.2\,MHz or $\sim$27\,km\,s$^{-1}$).\footnote{In TDM mode, the usable bandwidth is also $\sim 7.5$\,GHz after excluding the edges of the spectral windows with high noise.}  The total number of fields is 243 after excluding the data sets of the two [C\,{\sevensize II}]-detected SMGs at $z=4.4$ \citep{2012MNRAS.427.1066S}.  The total on-source exposure time of the data is 19.2 hours.  The primary beam size is FWHM$\sim 18''$ and the synthesised beam size ranges from $0''.4-3''.8$.  We note that, in our procedure, we potentially miss spatially extended [C\,{\sevensize II}] emitter whose size is much larger than the synthesised beam size.

\begin{figure*}
\includegraphics[width=180mm]{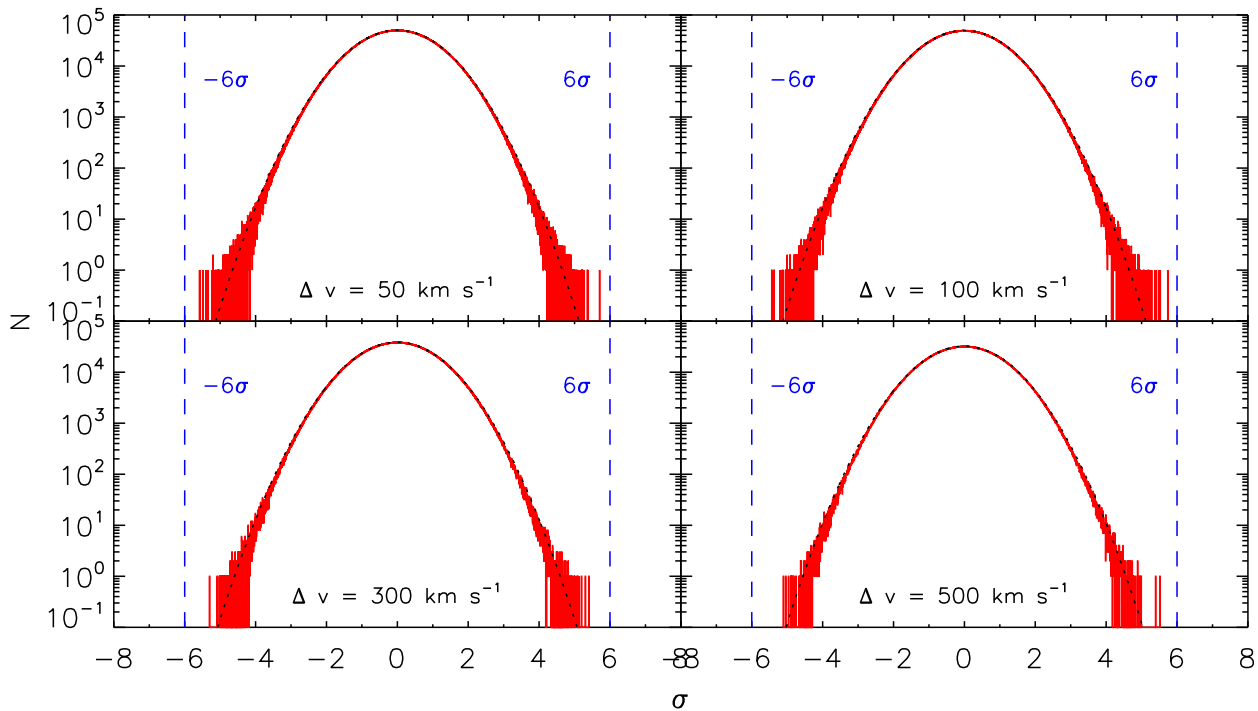}
  \caption{Flux distribution of the data cubes with four different spectral resolutions (50, 100, 300, and 500\,km\,s$^{-1}$) within the primary beam before primary beam correction.  The flux of each channel is normalised by using the rms.  The dotted curves show a Gaussian function.  The blue dashed vertical lines indicate $-6$-$\sigma$ and 6-$\sigma$, which is used as a detection threshold of line emitters.}
\end{figure*}

We used the ALMA data reduction package {\sc CASA}\footnote{http://casa.nrao.edu/} \citep{2007ASPC..376..127M} for continuum subtraction and imaging.  We used the calibrated data product provided by the archive, without any additional re-calibration of the data.  In order to ensure that the calibrations were correctly applied to the data products, we used {\sc clean} to image the phase calibrator from each of the datasets.  We confirmed that the phase calibrator is detected as a spatially un-resolved source at the phase centre and the derived flux density is consistent with the flux densities listed in the ALMA calibrator database.  We used {\sc uvcontsub} to subtract the continuum from the visibilities using line-free parts of the original targets with the multiple spectral windows.  The continuum subtraction procedure is necessary to subtract possible side lobes of a bright continuum source and to avoid to detect positive noise peaks on the bright continuum source as line emitters.   We used {\sc clean} to construct data cubes with natural weighting to maximise the sensitivity.  We adopted a pixel size of $0''.2$, which is $ \times 2$ smaller than the smallest synthesised beam size of $0''.4$ in the data.

In order to determine the binning size of the data cubes, we checked the [C\,{\sevensize II}] velocity range of high-$z$ galaxies/QSOs in literature.  Figure\,1 show a distribution of the [C\,{\sevensize II}] velocity width (FWHM) of 27 [C\,{\sevensize II}]-detected galaxies/QSOs at $z>4$ \citep{2011ApJ...740...63C, 2010A&A...519L...1W, 2012ApJ...752L..30W, 2012MNRAS.427.1066S, 2012A&A...543A.114G, 2006ApJ...645L..97I, 2013ApJ...763..120C, 2011A&A...530L...8D, 2014arXiv1404.2295D, 2014A&A...562A..35N, 2014ApJ...783...59R, 2013ApJ...773...44W, 2013Natur.496..329R, 2014arXiv1404.7159R, 2005A&A...440L..51M, 2009A&A...500L...1M,  2012MNRAS.425L..66M, 2009Natur.457..699W, 2013ApJ...770...13W, 2012ApJ...751L..25V}.  The [C\,{\sevensize II}] velocity width ranges from $\sim 50- 700$\,km\,s$^{-1}$ for a [C\,{\sevensize II}] luminosity range of $2 \times 10^8 L_{\odot} < L_{\rm [C\,{\sevensize II}]} < 2 \times 10^{10} L_{\odot}$.  The median [C\,{\sevensize II}] velocity width is FWHM\,$\sim$\,300\,km\,s$^{-1}$.  The 90 per cent of the sample are distributed in the range of $50-500$\,km\,s$^{-1}$.  Therefore, we made data cubes with four different spectral resolutions (50, 100, 300, and 500\,km\,s$^{-1}$) to search for line emitters in this velocity range.

Figure\,2 shows flux distributions of the data cubes within the primary beam before primary beam correction.  The flux is normalised by using the rms of each binned channel.  The blue dashed vertical lines indicate $-6$-$\sigma$ and 6-$\sigma$, which is used as a detection threshold of line emitters.  The overall shapes of the normalised flux distributions are well fitted with a Gaussian (The dotted curves).  However, the negative tails extend to larger than the 5-$\sigma$.  Therefore we set a threshold of 6-$\sigma$ for our line detection conservatively.  In this search, we could not detect any line emitters above the 6-$\sigma$ significance level.  We note that we excluded data with emission lines from the original targets and channels with very high noise, often seen at the edge of spectral windows.

We tested the reliability of our line emitter search as follows. Based on the data re-calibrated by \citet{2013ApJ...768...91H}, \citet{2012MNRAS.427.1066S} detected the $z=4.4$ [C\,{\sevensize II}] emission lines at 7.0-$\sigma$ from ALESS\,65.1 with $L_{\rm [C\,{\sevensize II}]} = 3.2 \pm 0.4 \times 10^{9} L_{\odot}$ and at 5.3-$\sigma$ from ALESS\,61.1 with $L_{\rm [C\,{\sevensize II}]} = 1.5 \pm 0.3 \times 10^{9} L_{\odot}$, respectively.  In our procedure, based on the data without re-calibration, we detected the emission line at 6.2-$\sigma$ from ALESS\,65.1 with $L_{\rm [C\,{\sevensize II}]} = 2.9 \pm 0.5 \times 10^{9} L_{\odot}$ and merginally detected the emission line at 3.8-$\sigma$ from ALESS\,61.1 with $L_{\rm [C\,{\sevensize II}]} = 1.5 \pm 0.4 \times 10^{9} L_{\odot}$.  Although the line luminosities measured in the two different procedures agree well, the significances in our procedure are somewhat less than those in \citet{2012MNRAS.427.1066S}.  These differences may come from different methods of data calibration and reduction.  This test demonstrated that our procedure can detect [C\,{\sevensize II}] emitters if they exist in the data cubes.

\begin{table*}
 \centering
 \begin{minipage}{160mm}
  \caption{Constraints for [C\,{\sevensize II}] luminosity function at $z\sim4.5$}
  \begin{tabular}{@{}ccccccccc@{}}
  \hline
 & \multicolumn{2}{c}{$\Delta v=50$\,km\,s$^{-1}$} &  \multicolumn{2}{c}{$\Delta v=100$\,km\,s$^{-1}$} &  \multicolumn{2}{c}{$\Delta v=300$\,km\,s$^{-1}$} &  \multicolumn{2}{c}{$\Delta v=500$\,km\,s$^{-1}$}\\
  \hline
  $L_{[\rm C\,{\sevensize II}]}$ & $V_{\rm survey}^a$ & $N^b$& $V_{\rm survey}^a$ & $N^b$  & $V_{\rm survey}^a$ & $N^b$ & $V_{\rm survey}^a$ & $N^b$ \\
   ($10^9 L_{\odot}$) & (cMpc$^3$) &  (cMpc$^{-3}$) &  (cMpc$^3$) &  (cMpc$^{-3}$) &  (cMpc$^3$) &  (cMpc$^{-3}$) &  (cMpc$^3$) &  (cMpc$^{-3}$)\\
 \hline
 $0.13$ & $24$ & $<7.8 \times 10^{-2}$ & $12$ & $<1.6 \times 10^{-1}$ & $0$ & --- & $0$ & --- \\
 $0.25$ & $97$ & $<1.9 \times 10^{-2}$ & $42$ & $<4.3 \times 10^{-2}$ & $12$ & $<1.5 \times 10^{-1}$ & $6.1$ & $<3.0 \times 10^{-1}$\\
 $0.5$ & $7.1 \times 10^{2}$ & $<2.6 \times 10^{-3}$ & $1.5 \times 10^{2}$ & $<1.2 \times 10^{-2}$ & $57$ & $<3.2 \times 10^{-2}$ & $31$ & $<6.0 \times 10^{-2}$\\
 $1$ & $3.5 \times 10^{3}$ & $<5.3 \times 10^{-4}$ & $1.4 \times 10^{3}$ & $<1.3 \times 10^{-3}$ & $2.8 \times 10^{2}$ & $<6.7 \times 10^{-3}$ & $1.0 \times 10^{2}$ & $<1.8 \times 10^{-2}$\\
 $2$ & $7.7 \times 10^{3}$ & $<2.4 \times 10^{-4}$ & $4.9 \times 10^{3}$ & $<3.8 \times 10^{-4}$ & $1.7 \times 10^{3}$ & $<1.1 \times 10^{-3}$ & $7.9 \times 10^{2}$ & $<2.3 \times 10^{-3}$\\
 $4$ & $1.1 \times 10^{4}$ & $<1.7 \times 10^{-4}$ & $8.4 \times 10^{3}$ & $<2.2 \times 10^{-4}$ & $4.7 \times 10^{3}$ & $<3.9 \times 10^{-4}$ & $3.0 \times 10^{3}$ & $<6.1 \times 10^{-4}$\\
 $8$ & $1.3 \times 10^{4}$ & $<1.5 \times 10^{-4}$ & $1.1 \times 10^{4}$ & $<1.7 \times 10^{-4}$ & $7.4 \times 10^{3}$ & $<2.5 \times 10^{-4}$ & $5.5 \times 10^{3}$ & $<3.4 \times 10^{-4}$\\
\hline
\end{tabular}
$^a$The survey volumes are calculated after the primary beam correction.\\
$^b$1$\sigma$ upper limits from non-detection \citep{1986ApJ...303..336G}.\\
\end{minipage}
\end{table*}

\begin{figure*}
\includegraphics[width=170mm]{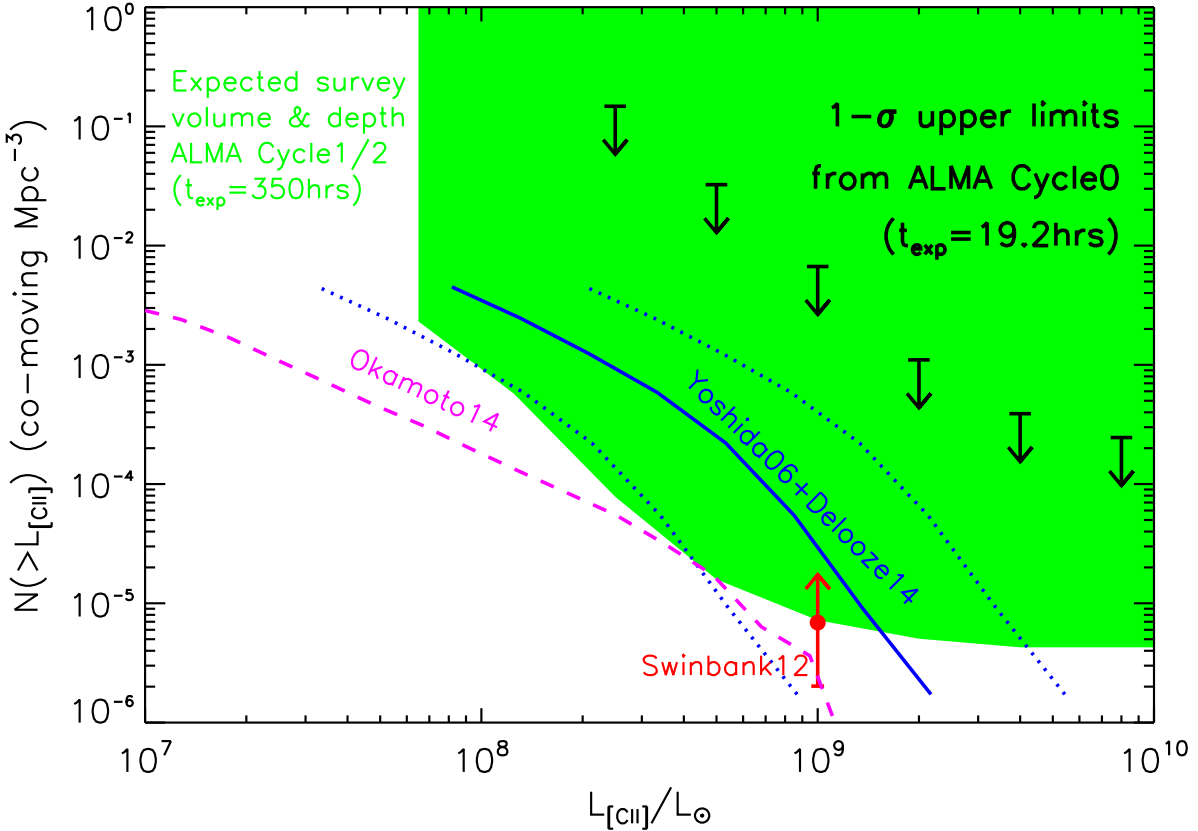}
  \caption{ALMA constraints for the [C\,{\sevensize II}] luminosity function at $z\sim 4.5$.  The black arrows show 1-$\sigma$ upper limits from non-detection of line-emitters in data cubes with $300$\,km\,s$^{-1}$ spectral resolution.  The red arrow indicates a lower limit at $z=4.44$ from \citet{2012MNRAS.427.1066S}.  The solid blue curve represents the [C\,{\sevensize II}] luminosity function expected from the $z \sim 4$ UV luminosity function \citep{2006ApJ...653..988Y} and the SFR/L$_{\rm [C\,{\sevensize II}]}$ calibration for high-$z$ galaxies \citep{2014arXiv1402.4075D}.  The dotted curves show the $\pm 1$-$\sigma$ uncertainties (0.4\,dex) of the $SFR/L_{\rm [C\,{\sevensize II}]}$ calibration.  We also adopted the dust attenuation in UV, $A_{UV}=1.0$ at $z \sim 4$, by \citet{2013A&A...554A..70B} and the relationship between UV luminosity density ($L_{\rm UV}$) and star-formation rate (SFR) by \citet{2012ARA&A..50..531K}.  The dashed curve is the [C\,{\sevensize II}] luminosity function predicted from a numerical simulation \citep{2014PASJ...66...70O} with the same empirical SFR/L$_{\rm [C\,{\sevensize II}]}$ calibration. The current upper limits are still at least $2$ orders of magnitude larger than the expected [C\,{\sevensize II}] luminosity functions.  The green area indicates the survey volume and depth expected from ALMA Cycle\,1 and 2 archival data.}
\end{figure*}

The non-detection of new line emitters provides upper limits for the [C\,{\sevensize II}] luminosity function at $z\sim 4.5$.  We listed the survey depths, survey volumes, and 1-$\sigma$ upper limits in Table\,3.  We calculated the survey volumes as a function of survey depth.\footnote{Data cubes with higher spectral resolutions have higher sensitivities for narrower emission lines and therefore have larger survey volumes for a given line luminosity limit.}  We took into account the primary beam correction for the calculation because the line sensitivity is not homogeneous even in the same channel.  We assume that line emitters are spatially unresolved in the data cubes.  The 1-$\sigma$ confidence upper limits on the space densities of [C\,{\sevensize II}] emitters are calculated using Poisson statistics by \citet{1986ApJ...303..336G}.  We plot the upper limits for the $\Delta v = 300$\,km\,s$^{-1}$ case as arrows in Figure\,3.  We note that, in addition to [C\,{\sevensize II}], [O\,{\sevensize III}]\,88\,$\mu$m, [N\,{\sevensize II}]\,122\,$\mu$m, [O\,{\sevensize I}]\,145\,$\mu$m, and [N\,{\sevensize II}]\,205\,$\mu$m at high redshift ($z\ge3$), and high-J CO at lower redshift (i.e. $z\ge 0.4$ for $J\ge3$) could be observed in ALMA band\,7 \citep[e.g.][]{2012MNRAS.427.1066S, 2013arXiv1301.0371C, 2014arXiv1403.4360O}.  However, the {\it upper} limits for the [C\,{\sevensize II}] luminosity function are not affected from the other possible line emitters at different redshifts.

\section[]{Discussions and Summary}

For comparison, we plot the [C\,{\sevensize II}] luminosity function expected from the $z\sim 4$ UV luminosity function from \citet{2006ApJ...653..988Y} in Figure\,3. We adopted a dust attenuation in the UV, $A_{UV}=1$ at $z\sim 4$ \citep{2013A&A...554A..70B}, the relationship between UV luminosity density ($L_{\rm UV}$) and star-formation rate (SFR), SFR(M$_{\sun}$\,yr$^{-1}$) $= 1.2 \times 10^{28} L_{\rm UV}$ (ergs\,s$^{-1}$\,Hz$^{-1}$) by \citet{2012ARA&A..50..531K}, $SFR/L_{\rm [C\,{\sevensize II}]}$ calibration for high-$z$ galaxies, log$SFR$(M$_{\sun}$\,yr$^{-1}$) $=  -8.52 + 1.18\times$log$L_{\rm [C\,{\sevensize II}]} (L_{\odot})$, by \citet[]{2014arXiv1402.4075D}.  We also compared with the [C\,{\sevensize II}] luminosity function at $z=4$ predicted from a numerical simulation \citep{2014PASJ...66...70O} with the same empirical $SFR/L_{\rm [C\,{\sevensize II}]}$ calibration.  The current upper limits are at least $2$ orders of magnitude larger than the expected [C\,{\sevensize II}] luminosity functions.  Given the depth and sensitivity improvements with ALMA in Cycle 1 and 2, we expect to be able to detect [C II] emitters in blind searches as ALMA Patchy Deep Survey (see the green region in Figure\,3).

If we adopt a detection threshold of 5.5-$\sigma$, we could detect 6 emitter candidates.  If these sources are real, we would have a number density of [C\,{\sevensize II}] emitters of $\sim 2 \times 10^{-2}$\,Mpc$^{-3}$ for $L_{\rm [C\,{\sevensize II}]}> 5 \times 10^8$\,L$_{\odot}$.  In this case, the [C\,{\sevensize II}] luminosity function (or the SFR density) at $z>4$ would be $\sim 2$ orders of magnitude higher than those expected from UV observations.  However, the negative tail of the noise distribution also extends to -5.5-$\sigma$ and the number of negative sources is 2 below -5.5-$\sigma$.  If we consider Poisson statistics, the 1-$\sigma$ uncertainty of the false positive is $2.0^{+2.6}_{-1.3}$.  The number excess of the "detected" sources is only $\sim 1.5$-$\sigma$ compared to the false positive.  Therefore, the 5.5-$\sigma$ detection threshold may be too aggressive to detect real source.

In summary, we carried out a blind search for [C\,{\sevensize II}] emitters at $z\sim 4.5$ using a part of ALMA Cycle\,0 archival extra-galactic data.  In the continuum-subtracted data cubes with spectral resolutions of $\sim$50-500\,km\,s$^{-1}$, we could not detect any new line emitters above a 6-$\sigma$ significance level.  However, we obtained the first upper limits to the [C\,{\sevensize II}] luminosity function at $z\sim 4.5$ over $L_{\rm [C\,{\sevensize II}]} \sim 10^8 - 10^{10} L_{\odot}$ or star formation rate, $SFR \sim 10-1000$~M$_{^\odot}$yr$^{-1}$.  These limits are at least 2 orders of magnitude larger than the [C\,{\sevensize II}] luminosity functions expected from UV observations or numerical simulations.  Future ALMA Patchy Deep Survey using archival data will be able to detect [C\,{\sevensize II}] emitters in the blind searches and to study the cosmic star-formation rate density at $z\ge 4$.

\section*{Acknowledgments}

We thank the referees for helpful comments which significantly improved the content and presentation of this paper.  We also thank Natsuki Hayatsu, Naoki Yoshida, Ian Smail, Hiroshi Nagai, Daniel Espada, Ken Tatematsu, Soh Ikarashi and Mark Lacy for their help and discussions.  This paper makes use of the following ALMA data:ADS/JAO.ALMA\#2011.0.00020.S, \#2011.0.00039.S,  \#2011.0.00097.S, \#2011.0.00101.S, \#2011.0.00108.S, \#2011.0.00208.S, \#2011.0.00294.S, and \#2011.0.00467.S.  ALMA is a partnership of ESO (representing its member states), NSF (USA) and NINS (Japan), together with NRC (Canada) and NSC and ASIAA (Taiwan), in cooperation with the Republic of Chile.  The Joint ALMA Observatory is operated by ESO, AUI/NRAO and NAOJ.  Data analysis were carried out on common use data analysis computer system at ADC/NAOJ.  YM acknowledges support from JSPS KAKENHI Grant Number 20647268.

\label{lastpage}

\end{document}